\documentclass[%
 reprint,
superscriptaddress, amsmath,amssymb, pra]{revtex4-1}

\usepackage{graphicx}
\usepackage{dcolumn}
\usepackage{bm}
\usepackage{hyperref}
\usepackage{physics}
\usepackage[utf8]{inputenc}
\usepackage[T1]{fontenc}

\usepackage{xcolor}

\def\equationautorefname~#1\null{%
  Eq.~(#1)\null
}

\def\figureautorefname~#1\null{%
  Fig.~#1\null
}

\begin{document}

\title{Long-range exchange interaction between spin qubits mediated by a superconducting link at finite magnetic field
}

\author{L. Gonz\'alez Rosado}
\affiliation{JARA Institute  for  Quantum  Information,  RWTH  Aachen  University,  52056  Aachen, Germany}
\affiliation{JARA Institute for Quantum Information (PGI-11), Forschungszentrum J\"ulich, 52425 J\"ulich, Germany}

\author{F. Hassler}
\affiliation{JARA Institute  for  Quantum  Information,  RWTH  Aachen  University,  52056  Aachen, Germany}
\author{G. Catelani}
\affiliation{JARA Institute for Quantum Information (PGI-11), Forschungszentrum J\"ulich, 52425 J\"ulich, Germany}

\date{\today}

\begin{abstract}
Solid state spin qubits are promising candidates for the realization of a quantum computer due to their long coherence times and easy electrical manipulation. However, spin-spin interactions, which are needed for entangling gates, have only limited range as they generally rely on tunneling between neighboring quantum dots. This severely constrains scalability. Proposals to extend the interaction range generally focus on coherent electron transport between dots or on extending the coupling range. Here, we study a setup where such an extension is obtained by using a superconductor as a quantum mediator. Because of its gap, the superconductor effectively acts as a long tunnel barrier. We analyze the impact of spin-orbit (SO) coupling, external magnetic fields, and the geometry of the superconductor. We show that while spin non-conserving tunneling between the dots and the superconductor due to SO coupling does not affect the exchange interaction, strong SO scattering in the superconducting bulk is detrimental. Moreover, we find that the addition of an external magnetic field decreases the strength of the exchange interaction. Fortunately, the geometry of the superconducting link offers a lot of room to optimize the interaction range, with gains of over an order of magnitude from a 2D film to a quasi-1D strip. We estimate that for superconductors with weak SO coupling (\textit{e.g.}, aluminum) exchange rates of up to 100\,MHz over a micron-scale range can be achieved with this setup in the presence of magnetic fields of the order of 100\,mT. 
\end{abstract}
\maketitle
\section{\label{sec:intro}INTRODUCTION}

The field of quantum computation advances rapidly, with the first quantum computers already outperforming classical computers for certain tasks~\cite{AABetal}. However, the first prototypes consist merely of a few dozen qubits, and the idea of an universal quantum computer made of thousands of qubits remains a distant goal for now. While progress is being made in a variety of qubit architectures~\cite{DS,BXNOetal,GALHC}, scalability remains a common challenge for all of them. One of the most promising qubit architecture are semiconductor-based spin qubits. Their main advantages over competing alternatives are the long coherence times and easy qubit manipulation, together with the straightforward production of quantum dots (QDs) by standard lithographic techniques~\cite{LD,KL,VBCDetal}. In these setups, two-qubit gates are conventionally realized by exchange interactions. However, such interactions are short-ranged, which heavily constrains the spatial distance between QDs and impedes scalability. As a result, engineering long-range interaction between spin-qubits in QDs has been an active field of research in recent years~\cite{MKFBetal, TELWetal, BSFRetal,NHHLetal,WD,LPFDetal}. 

A promising approach to extend the range of the interaction in QD-based spin qubits is the use of a quantum mediator. Examples of such systems include long-range interaction mediated by a third quantum dot~\cite{BGGSetal,BFRWetal,Dzurak}, floating metallic gates~\cite{TDTWetal,SKL}, quantum Hall edge states~\cite{Loss,Doherty}, and superconductors~\cite{Mi156,Flensberg,vanWoe,SPC}. Here, we focus on the latter example, basing our work on a proposal where the exchange coupling between two quantum dots is mediated by a thin superconducting film that is tunnel-coupled to the dots~\cite{HCB}. An effective coupling between the dots is mediated by virtual transitions in and out of the superconducting film. In Ref.~\cite{HCB}, it was estimated that exchange interaction strengths of the order of $10-100\,$MHz over length scales of a few micrometers for a two-dimensional superconducting film can be achieved in this setup. In this paper, we build on the previous proposal and consider three additional effects of experimental relevance: the possibility of spin non-conserving tunneling from the dots to the superconductor due to spin-orbit coupling at the interface, the addition of an external magnetic field, and the role of the geometry of the superconducting film, in particular, the 2D to 1D crossover. 

We show that the addition of SO induced spin non-conserving tunneling between the dots and the superconductor is equivalent to a controlled spin rotation. In the absence of an external magnetic field, this can be taken into account by a fixed rotation of the spin quantization axis and does not affect the strength of the exchange interaction. On the other hand, SO scattering in the superconducting bulk leads, due to disorder averaging over different paths with variations in the spin-rotation, to a decrease in the effective coherence length of the exchange interaction for distances larger than spin-orbit length $l_{\text{so}}$. We study the effect of an external magnetic field which is commonly used for qubit manipulation and read-out~\cite{WWPDetal,TNYNT,SDHBetal}. Assuming that the field is weak enough, such that the superconducting gap is not affected, and oriented parallel to the thin superconducting film, it creates a Zeeman splitting in both the dots and the superconductor~\cite{Fulde}. As the degeneracy of the energy levels is broken due to the Zeeman splitting, the energy of the electrons in the virtual intermediate state varies for the different processes (\textit{i.e.}, different initial spins, and spin conserving versus non-conserving tunneling). In order to ensure that all processes remain entirely virtual such that the electron cannot leak to the quasiparticle states above the gap, a retuning of the energy levels of the quantum dot is necessary. This ultimately leads to a reduction of the effective exchange-coupling between the two quantum dots. We also investigate the influence of the geometry of the superconductor, in particular, the crossover from a 2D to a quasi-1D configuration. In contrast to SO coupling and the external magnetic field, reducing the effective dimensionality of the superconductor has the potential to increase the exchange interaction by over an order of magnitude in comparison to the infinite two-dimensional case discussed in Ref.~\cite{HCB}. 
In particular, we show that for the specific case where the superconducting film is made of aluminum, a material with very weak spin-orbit scattering~\cite{CWA}, and for magnetic fields of the order of $100\,\text{mT}$, an exchange interaction of the order of $100\,\text{MHz}$ can be obtained. We want to point out that the decrease due to the external magnetic field and SO scattering can be more than compensated by reducing the width of the superconducting link. 

The outline of the paper is as follows. In \autoref{sec:setup} we introduce the setup. Section \ref{sec:exchange} contains the main results of our manuscript. The section is divided in three subsections: the first subsection discusses the effects on the exchange interaction of spin-orbit coupling, the second subsection considers the full two-dimensional model, with both spin-orbit effects present as well as an external field, and the third subsection tackles the impact of the geometry of the superconductor. Finally, in \autoref{sec:conclusiones} we summarize our findings and give estimates for the achievable ranges of coupling strength for a superconducting coupler made of aluminum.

\section{\label{sec:setup}setup}

\begin{figure}[tb]
 \includegraphics[width=\linewidth]{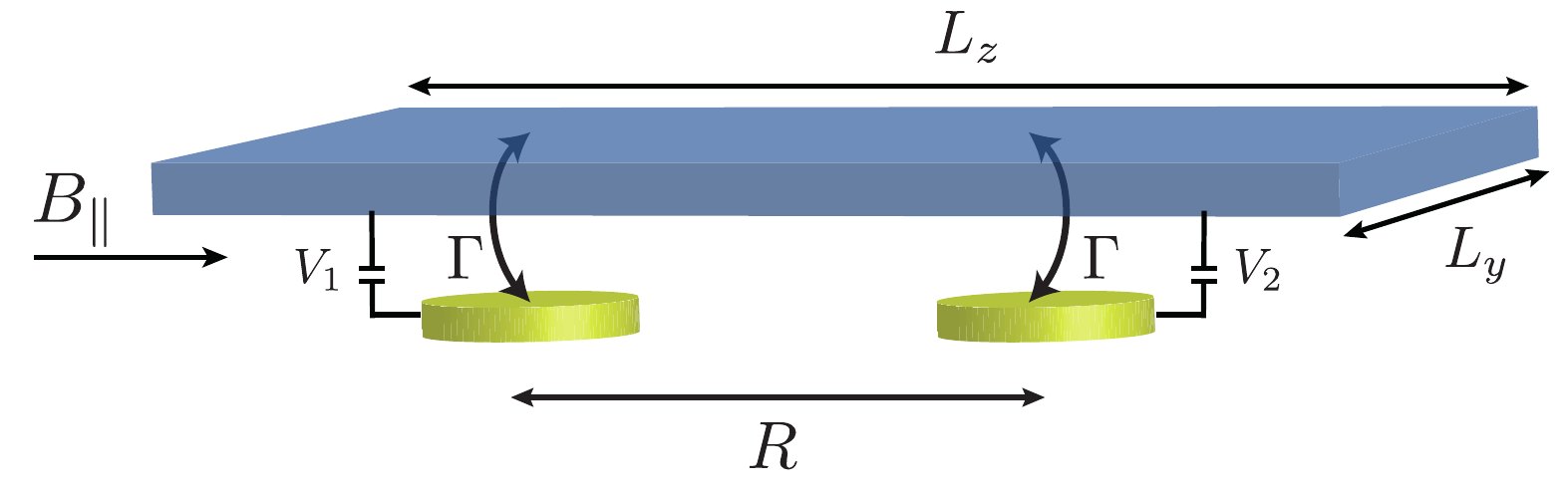}
  \caption{
Schematic representation of the setup: two quantum dots at gate voltages $V_1$ and $V_2$ are separated by a distance $R$ along the $z$ direction. They are tunnel-coupled (with rate $\Gamma$) to a conventional superconducting 2D film of thickness $L_x$ smaller than the superconducting coherence length. We analyze the situation where a weak magnetic field $B_{\parallel}$ is applied to the system parallel to the superconducting film (\textit{i.e.}, in the $yz$-plane), which induces a Zeeman splitting in both the dots and the superconductor.}
  \label{fig:setup}
\end{figure}%

We study the exchange interaction between electron spins in two quantum dots coupled via a thin superconducting film (see \autoref{fig:setup}). We extend the results of Ref.~\cite{HCB} by including the effects of 
an external magnetic field $B_{\parallel}$ parallel to the superconducting film, spin non-conserving tunneling events between the dots and the superconductor due to SO coupling, and the geometry of the superconducting film.
The setup is modeled by the Hamiltonian 
\begin{equation}
    H=H_D+H_{\text{BCS}}+H_T+H_Z
\end{equation}
where $H_D$ is the Hamiltonian of the quantum dots, $H_{\text{BCS}}$ of the superconducting film, $H_T$ describes the tunneling between the dots and the superconductor, and $H_Z$ the external magnetic field. We discuss these terms in detail in the
following, setting $\hbar= 1$ throughout the text.

The dots are described by $H_D=H_1+H_2$, where $H_1$ and $H_2$ model the first (left) and second (right) dot, respectively. The Hamiltonians of the individual quantum dots are given by
\begin{equation}
    H_j=\epsilon_j n_j +\frac{1}{2}U_j n_j(n_j-1)
    \label{eq:hdot}
\end{equation}
with the number operator $n_j=\sum_{\sigma} d^\dag_{j\sigma}d_{j\sigma}$ where $d^\dag_{j\sigma}$ and $d_{j\sigma}$ the creation and annihilation operators for an electron of spin $\sigma$ in the dot $j$. The energy of the lowest occupation level in the dot, measured from the chemical potential of the superconductor $\mu$, is given by $\epsilon_j$, which we assume to be experimentally tunable by nearby gates. The term proportional to $U_j>0$ describes the repulsive Coulomb interaction between electrons in the dot. 

\begin{figure}[tb]
  \includegraphics[width=\linewidth]{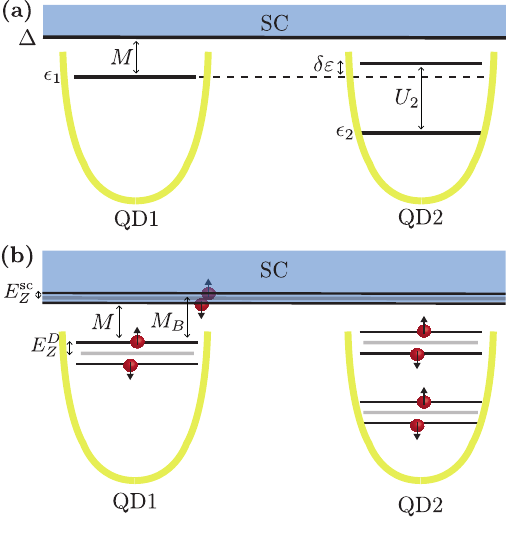}
    \caption{
(a) Level scheme of the quantum dots in the absence of an external magnetic field. The energy levels of the quantum dots are depicted inside the yellow parabolas, which are placed at the position of the first (QD1) and second (QD2) quantum dot. The blue area represents the superconducting density of states. (b) The external magnetic field creates a Zeeman splitting that leads to distinct energy level schemes for the spin-up and the spin-down states ($g^D,g^{\text{sc}}>0$). We denote with $M$ the minimum energy difference between an electron in the first quantum dot and in the superconductor. Note that the minimum energy difference for spin-conserving tunneling is different, given by $M_B> M$.}
    \label{fig:energies}
\end{figure}

In the following, we consider the situation where the states $\ket{\sigma,\sigma'}$ (one electron with spin $\sigma$ in the first dot, and an electron with spin $\sigma'$ in the second one) and $\ket{0,\uparrow\downarrow}$ (both electrons in the right dot in a spin-single configuration) are close in energy. In particular, we set the energy difference $\delta \varepsilon = \varepsilon_{0,2}-\varepsilon_{1,1} >0$ with $\varepsilon_{1,1}=\epsilon_1+\epsilon_2$ and $\varepsilon_{0,2}=2\epsilon_2+U_2$, much smaller than the energy spacing in the dots. A schematic representation of the energy levels can be found in \autoref{fig:energies}(a). Note that by choosing these two states to be very close in energy, the states $\ket{\uparrow\downarrow,0}$ and $\ket{\uparrow,\downarrow}$ have a large energy offset. This breaks the inversion symmetry of the system and allows to only consider an electron virtually travelling from the first to the second dot, and not the other way around.

The superconducting film, assumed to be of $s$-wave pairing, is modeled by the BCS mean-field Hamiltonian~\cite{BCS}
\begin{equation}
    H_{\text{BCS}}=\sum_{\bm{k},\sigma}\epsilon_kc^\dag_{\bm{k}\sigma}c_{\bm{k}\sigma}-\Delta \sum_{\bm{k}} c^\dag_{\bm{k}\uparrow}c^\dag_{-\bm{k}\downarrow} +\text{H.c.}
    \label{eq:hbcs}
\end{equation} where $c^\dag_{\bm{k}\sigma}$ and $c_{\bm{k}\sigma}$ denote the creation and annihilation of an electron of momentum $\bm{k}$ and spin $\sigma$ in the superconductor; here, $\Delta >0$ is the energy gap of the superconductor, and $\epsilon_k$ is the electron energy measured with respect to the chemical potential of the superconductor. We assume a dirty superconductor where disorder leads to diffusive motion due to elastic scattering. In particular, we expect scattering at the boundaries of the superconductor to be important, limiting the mean-free path to $L_x$. We model the disorder by a random potential $V(\bm r)$ with Gaussian statistics with $\overline{V(\bm{r})}=0$ and $\overline{V(\bm{r})V(\bm{r}')} =\gamma_e \delta^{(d)}(\bm{r}-\bm{r}')$, where the overline denotes the average over different disorder configurations. The disorder parameter $\gamma_e$ is related to the mean free-path $\ell_e$ and the density of states per spin in the normal state $\rho_0$ via $\gamma_e=v_F/2\pi \rho_0 \ell_e$ ($v_F$ denotes the Fermi velocity). We average the dynamics of the electrons in the superconductor over disorder with the use of diagrammatic techniques (see \cite{GRHC} and reference therein for details).

We model the coupling between the dots and the superconductor by the most general time-reversal invariant tunneling Hamiltonian~\cite{DN}, which we divide in two terms as $H_T=H_T^0+H_T^F$, where $H_T^0$ is spin-conserving tunneling and $H_T^F$ spin non-conserving. In particular, we find 
\begin{align}\label{eq:htunnel}
H_{T}^0&=-\frac{1}{\sqrt{L_yL_z}}\sum_{\bm{k},\sigma}[t_1c_{\bm{k},\sigma}^{\dag}d_{1\sigma}+e^{-ik_zR}t_2c_{\bm{k},\sigma}^{\dag}d_{2\sigma}]+\text{H.c.}\nonumber\\
H_{T}^{\text{F}}&=-\frac{1}{\sqrt{L_yL_z}}\sum_{\bm{k}}[t_{f1}c_{\bm{k}\uparrow}^{\dag}d_{1\downarrow}-t_{f1}^* c_{\bm{k}\downarrow}^{\dag}d_{1\uparrow}\nonumber\\&\qquad+e^{-ik_zR}(t_{f2}c_{\bm{k}\uparrow}^{\dag}d_{2\downarrow}-t_{f2}^*c_{\bm{k}\downarrow}^{\dag}d_{2\uparrow})]+\text{H.c.},
\end{align}%
where $t_1,t_2 >0$, $t_{f1},t_{f2} \in \mathbb{C}$, and $R$ is the distance between the two dots (which we take to be along the $z$-axis). For concreteness, we have assumed that tunneling is into an effectively two-dimensional superconductor, so from now on the density of states $\rho_0$ should be understood as the appropriate 2D one. We also introduce the total tunneling rate $\Gamma=2\pi t^2\rho_0$, where $t$ is the total tunneling amplitude $t=(t_1^2+|t_{f1}|^2)^{1/2}=(t_2^2+|t_{f2}|^2)^{1/2}$ taken to be the same in both dots for simplicity.

We assume that the magnetic field with magnitude $B$ is oriented parallel to the superconducting film. Moreover, it should be weak enough that the superconducting gap is not affected; that is, we assume $B \lesssim 0.5 B_{c\parallel}$, where $B_{c\parallel}$ is the parallel critical field. We include the effect of the magnetic field via a Zeeman splitting in both the superconductor as well as the quantum dots which takes the form
\begin{equation}
H_Z=E_Z^{\text{sc}}\sum_{\bm{k}}( c^\dag_{\bm{k}\uparrow}c_{\bm{k}\uparrow}\!-\!c^\dag_{\bm{k}\downarrow}c_{\bm{k}\downarrow})+E_Z^{D}\sum_{i}(d_{i\uparrow}^\dag d_{i\uparrow}\!-\!d_{i\downarrow}^\dag d_{i\downarrow}),
\label{eq:hz}
\end{equation}
 where the spin quantization axis is chosen along the field direction. Here, $E_Z^{\text{sc}}=\frac{1}{2}\mu_B g^{\text{sc}}B$ and $E_Z^{D}=\frac{1}{2}\mu_B g^{D}B$, with $\mu_B$ the Bohr magneton and $g^{\text{sc},D}$ the $g$-factors, which we assume to be the same in both quantum dots, but different in the superconductor.

In the next section, we calculate the exchange interaction in this model and in order to do so, we first revisit the results obtained in Ref.~\cite{HCB} for $H_T^F=H_Z=0$. 

\section{\label{sec:exchange}Exchange interaction}

\begin{figure}[tb]
  \includegraphics[width=\linewidth]{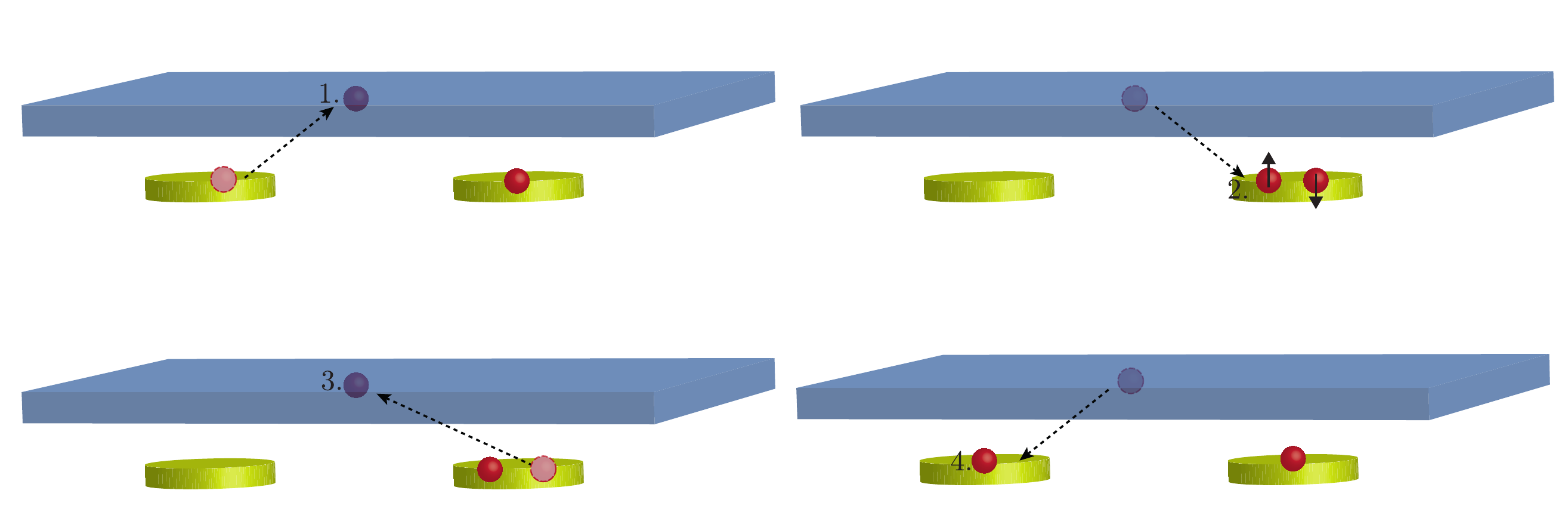}
    \caption{Fourth order perturbation theory leading to an exchange interaction: (top row) a second order process, involving a virtual intermediate state with an electron in the superconductor, couples the initial state with one electron in each dot to a state where both electrons reside in the same quantum dot (subject to Pauli's exclusion principle). The following second order process (bottom row) in the reverse direction brings the system back to its ground state and leads to an effective exchange interaction.}
    \label{fig:process}
\end{figure}

The exchange interaction stems from fourth order perturbation theory in the tunneling Hamiltonian (see \autoref{fig:process}). It takes the form $H_\text{ex}=\frac{1}{4}J\bm{\sigma}^1\cdot\bm{\sigma}^2$, where $J$ is the strength of the interaction, and $\bm{\sigma}^{1,2}$ the vectors of Pauli matrices for the first and second dot respectively. We define the dimensionless coupling parameter $\alpha$ by $J=\alpha\Gamma^2/\delta\varepsilon$. In the simplified case studied in Ref.~\cite{HCB}, where $H_Z=H_T^F=0$, it is shown that
\begin{equation}
        \alpha=\frac{1}{\Gamma^2}\overline{\left| \sum_{\bm{k},\sigma}\frac{\bra{0,\uparrow\downarrow}H_T\ket{\bm{k}\sigma;0,\bar{\sigma}}\bra{\bm{k}\sigma;0,\bar{\sigma}}H_T\ket{\sigma,\bar{\sigma}}}{\varepsilon_{1,1}-\varepsilon_{0,1}-E_k}\right|^2},
        \label{eq:alphapert}
\end{equation}
where $E_k=\sqrt{\epsilon_k^2+\Delta^2}$ denotes the quasiparticle spectrum of the superconductor, $\bar{\uparrow} = {\downarrow}$ and vice versa. Here, $\ket{\bm{k}\sigma;0,\sigma}$ denotes the intermediate state where the electron with spin $\sigma$ from the left dot is promoted to a quasiparticle with momentum $\bm k$ in the superconductor. 

The energies $\varepsilon_{1,1}$ and $\varepsilon_{0,1}$ play a key role in the exchange interaction through the denominator of \autoref{eq:alphapert}. Their difference should be minimized in order to increase the effective coupling between the dots. We define the minimum detuning $M$ as the minimum energy difference between the superconducting gap and the energy of an electron in the first dot (see \autoref{fig:energies}). Due to the broadening of the superconducting density of states, $M$ cannot be arbitrarily low but we can safely assume $M \ll \Delta$. In the absence of a magnetic field, this leads to detuning $\varepsilon_{1,1}-\varepsilon_{0,1}=\epsilon_1=\Delta-M$. 

The expression for $\alpha$ in \autoref{eq:alphapert} can be written in terms of the electronic Green's functions in a superconductor. In the two dimensional limit for $M\ll \Delta \ll \mu$, as shown in Ref.~\cite{HCB}, it assumes the form
\begin{equation}
\begin{split}
    \alpha&=\frac{1}{2\pi^2\rho_0^2}\overline{g(\Delta-M,R)g(\Delta-M,R)}\\
    &=\frac{\Delta}{2 \pi M k_F \ell_e} K_0(R/\xi_D)
    \end{split}
    \label{eq:alpha0}
\end{equation}
with the effective coherence length $\xi_D= D^{1/2}/(8\Delta M)^{1/4}$, where $D=v_F \ell_e/d$ is the diffusion constant and where the electronic Green's function is defined as
\begin{equation}
    g(E,R)=\int \frac{d^dk}{(2\pi)^d} \frac{(E+\epsilon_k)}{E^2-\Delta^2-\epsilon_k^2} e^{i\bm{k}\cdot \bm{R}}.
    \label{eq:greenf}
\end{equation}%
In 2D, the decay of the exchange interactions is controlled by the Macdonald function $K_0(x) = \int_0^\infty\!dt\, \cos[ x \sinh(t)]$.
Details on how to perform the average over disorder to obtain the result in \autoref{eq:alpha0} can be found in Appendix~\ref{appendix:angreen}.

\subsection{\label{sec:spinorbit}Spin-orbit effects}

Spin-orbit interaction can give rise to spin non-conserving tunneling events~\cite{DN} which we model by $H_T^F$. We show that the presence of such a spin non-conserving term can be absorbed into a fixed spin rotation. In the absence of a magnetic field, the exchange-interaction strength is therefore not affected by spin non-conserving tunneling. Note however the lowest energy state is not necessarily the singlet $\ket{S}=(\ket{\uparrow,\downarrow}-\ket{\downarrow,\uparrow})/\sqrt{2}$ any more, as the rotation in the quantization axis is in principle different for each of the dots. Still the resulting exchange interaction can be used to entangle the spin qubits.

The spin-rotation is explicitly given by the new operators
\begin{equation}
    \begin{split}
        d'_{j\uparrow}&=\cos(\varphi_j)d_{j\uparrow}+e^{i\alpha_j}\sin(\varphi_j)d_{j\downarrow}\\
        d'_{j\downarrow}&=\cos(\varphi_j)d_{j\downarrow}-e^{-i\alpha_j}\sin(\varphi_j)d_{j\uparrow}
    \end{split}
    \label{eq:dtilde}
\end{equation}
with $t_{fj}=|t_{fj}|e^{i\alpha_j}$ and $\varphi_j=\arctan(
|t_{fj}|/t_j)$. In terms of them, the tunneling Hamiltonian $H_T=H_T^0+H_T^F$ assumes the form
\begin{equation}
H_{T}=-\frac{t}{\sqrt{L_yL_z}}\sum_{\bm{k},\sigma}[ c_{\bm{k}\sigma}^{\dag} d'_{1\sigma}+e^{-ik_zR} c_{\bm{k}\sigma}^{\dag} d'_{2\sigma}]+\text{H.c}.
\end{equation}%
We see that after applying the rotation, $H_T$ is the same as in the absence of spin non-conserving tunneling. The same is true for $H_D$ but not for $H_Z$. This implies that in the absence of field, spin non-conserving tunneling simply contributes to the total tunneling probability. We discuss the case with magnetic field in the next section.

This result does not mean however, that spin-orbit coupling has no impact on the exchange interaction. We have so far only taken into account the effects of spin-orbit by including spin non-conserving terms in the tunneling between the dots and the superconductor. Spin-orbit coupling might also lead to spin rotations as the electron virtually travels through the superconducting film, which can be modelled by introducing spin-orbit scattering in the disorder potential. This process is different to the one modelled by $H_T^F$, as the path the electron takes in the superconductor is not fixed and neither is its spin rotation. Indeed, if a given disorder path between the first and second dot defines a given spin rotation for the electron, disorder averaging throughout the different paths will reduce the distinction between the different spin states, and as such, the exchange interaction. In particular, a new length scale $l_{\text{so}}=\sqrt{D\tau_{\text{so}}}$, where $\tau_{\text{so}}$ is the spin-orbit scattering time, will emerge~\cite{AM,GRHC}. The new effective coherence length $\xi_D^{\text{so}}$, with $(\xi_D^{\text{so}})^{-2}=\xi_D^{-2} +l_{\text{so}}^{-2}$, will determine the range of the exchange interaction as in \autoref{eq:alpha0}. As a result, the spin orbit coupling leads to the upper bound $\xi_D^{\text{so}} < l_{\text{so}}$ on the effective coherence length, but can be neglected for weak SO coupling when $l_{\text{so}} \gg \xi_D$.

\subsection{\label{sec:zeeman}External magnetic field}

In the previous subsection, we have studied the effects of SO coupling in our setup in the absence of an external magnetic field. We now discuss the effect of an external magnetic field parallel to the superconducting film as typically applied for spin-qubit operation. We work under the assumption that the magnetic field applied is weak enough that the superconducting gap of the coupler is not affected ($B \lesssim 0.5 B_{c\parallel}$). The magnetic field then induces a Zeeman splitting in the system in the form of \autoref{eq:hz}. We work in the limit $E_Z^D\gg t_1|t_{f1}|g(\Delta-M,0)$, such that the coupling between the states $\ket{\uparrow,\uparrow}$ and $\ket{\downarrow,\uparrow}$, as well as $\ket{\downarrow,\downarrow}$ and $\ket{\uparrow,\downarrow}$ due to spin non-conserving tunneling effects are small and can be neglected. Thus we focus for simplicity on the exchange subspace $\mathcal{E}$ spanned by the states $\mathcal{E}=\{\ket{\uparrow,\downarrow},\ket{\downarrow,\uparrow}\}$.

The effective Hamiltonian in $\mathcal{E}$ takes the form
\begin{equation}
    H_{\mathcal{E}}=\frac{J}{2} \tau_x+\beta \tau_z
    \label{eq:hqubit}
\end{equation}
where the $\tau_i$ matrices are the Pauli matrices acting on $\mathcal{E}$~\footnote{We do not include a contribution from $\tau_y$, corresponding to cross-terms in the exchange interaction in the form $\sigma_1^i \sigma_2^j$ with $i\neq j$, as they can be set to zero with a local basis rotation in the $xy$-plane.}. The energy splitting $\beta$ arises from the difference between the $g$-factors in the dots and in the superconductor and from spin non-conserving tunneling, which lead to a different effective magnetic field in the first dot due to second order perturbation terms in the tunneling Hamiltonian. It is calculated, along with $J$, which is given by fourth order corrections in the tunneling Hamiltonian, in Appendix \ref{appendix:heff}. The splitting $\beta$ [given in \autoref{eq:bz}] can be tuned by $\Delta_Z=E_Z^D-E_Z^{\text{sc}}$ and $E_{\text{tot}}=E_Z^D+E_Z^{\text{sc}}$ and is independent of $\delta\varepsilon$, whereas the exchange coupling $J=\Gamma^2 \alpha /\delta\varepsilon$ can be independently tuned by $\delta\varepsilon$.

Due to the Zeeman splitting, the energy level scheme is now more complex than in the absence of magnetic field (see \autoref{fig:energies}). In order to make sure that all processes remain virtual, we want to ensure that the detuning between the superconducting gap and the energy of an electron in the first dot does not reach values smaller than the minimum detuning $M$. Therefore, we  set $\epsilon_1=\Delta-M-E_{\text{tot}}$, assuming both $g$-factors to be positive for simplicity \footnote{If the $g$-factors have different sign, or if SO coupling is weak (such that the possible decay into the superconductor due to spin non-conserving tunneling as electrons reach energies higher than $\Delta-M$  could be neglected), the energy tuning $\epsilon_1=\Delta-M-|\Delta_Z|$ would give stronger exchange interaction.}. The dimensionless exchange coupling $\alpha$ is then given by (see Appendix~\ref{appendix:heff})
\begin{align}         \label{eq:alpha}
         \alpha&=\frac{2}{\Gamma^2}\Bigl[t_1^2t_2^2\overline{g(\Delta-M_B,R)g(\Delta-M_B-2\Delta_Z,R)}
         \nonumber\\&
       \quad+|t_{f1}|^2|t_{f2}|^2\overline{g(\Delta-M,R)g(\Delta-M-2E_{\text{tot}},R)}
        \nonumber\\&
       \quad+t_1t_2|t_{f1}||t_{f2}|\Bigl(\overline{g(\Delta-M_B,R)g(\Delta-M-2E_{\text{tot}},R)}
        \nonumber\\&
       \quad+\overline{g(\Delta-M,R)g(\Delta-M_B-2\Delta_Z,R)}\Bigr)\Bigr].
\end{align}
Here, $g(E,R)$ denotes the electronic Green's function in the superconductor [see \autoref{eq:greenf}] which is, for $E<\Delta$, entirely real, and $M_B=M+2E^{\text{sc}}_Z=M+E_{\text{tot}}-\Delta_Z$ [see also \autoref{fig:energies}(b)]. 

We take the spin non-conserving terms to give a negligible contribution to the total tunneling amplitude, $|t_{f1}|, |t_{f2}|\ll t$
~\cite{GG,Danon,MFRBetal}. As a result, \autoref{eq:alpha} can be approximated by
\begin{equation}
\begin{split}
    \alpha&\approx\frac{1}{2\pi^2 \rho_0^2} \overline{g(\Delta-M_B,R)g(\Delta-M_B-2\Delta_Z,R)}.
    \label{eq:alphatuned}
    \end{split}
\end{equation}

In Appendix~\ref{appendix:angreen}, we study the behavior of a disorder averaged pair of superconducting Green's functions at energies below the gap in a two dimensional superconductor. We can use these results to obtain an analytical approximation for $\alpha$ in two specific limits of interest:
\begin{equation}
    \alpha=\frac{\Delta K_0(R/\xi_B) }{2\pi k_F \ell_e }\begin{cases}
    M_B^{-1}, & M_B \gg 2\Delta_Z,\\
    (2\Delta_Z M_B)^{-1/2}, &  M_B \ll 2\Delta_Z.
    \end{cases}
    \label{eq:alphacases}
\end{equation}
The effective coherence length $\xi_B$ depends now on the external magnetic field via 
\begin{equation}
\frac{\xi_B}{\xi_D}=\begin{cases}
\Bigl(\frac{M}{M_B}\Bigr)^{1/4}, & M_B \gg 2\Delta_Z, \\
    \\ \Bigl(\frac{2M}{\Delta_Z}\Bigr)^{1/4}, &  M_B \ll 2\Delta_Z,
\end{cases}
    \label{eq:xis}
\end{equation}
where $\xi_D$ is the effective coherence length in the absence of an external magnetic field as defined after \autoref{eq:alpha0}.
Note that $K_0(x)$ can be approximated as $K_0(x) \simeq \sqrt{\pi/2 x}\, e^{-x}$ for $x\gg1$. Therefore, it is crucial that the interdot distance $R$ does not become much larger than $\xi_B$ in order to avoid an exponential suppression of the exchange interaction. Thus, we will focus on the regime $R \lesssim \xi_B$. In this limit, the previously introduced approximation for the Macdonald function for $x\gg 1$ is still valid, with an error of $15 \%$ for $x\approx 0.5$, where the behavior of $\alpha$ is dominated by the prefactor $\sqrt{\pi/2x}$ rather than by the exponential decay.

In the following, we analyze the increase in exchange interaction that can be obtained by decreasing the size of the superconducting film, and show that obtaining a factor of $10$ increase in $\alpha$ over the two-dimensional limit calculated thus far is possible by reducing the width of the superconductor $L_y$. We also show that this gain is comparable to the loss obtained from  considerably large Zeeman splittings of order $E_{\text{tot}}\approx 10M$.

\subsection{\label{sec:dimensionality}Dimensional crossover}

\begin{figure}[t]
  \includegraphics[width=\linewidth]{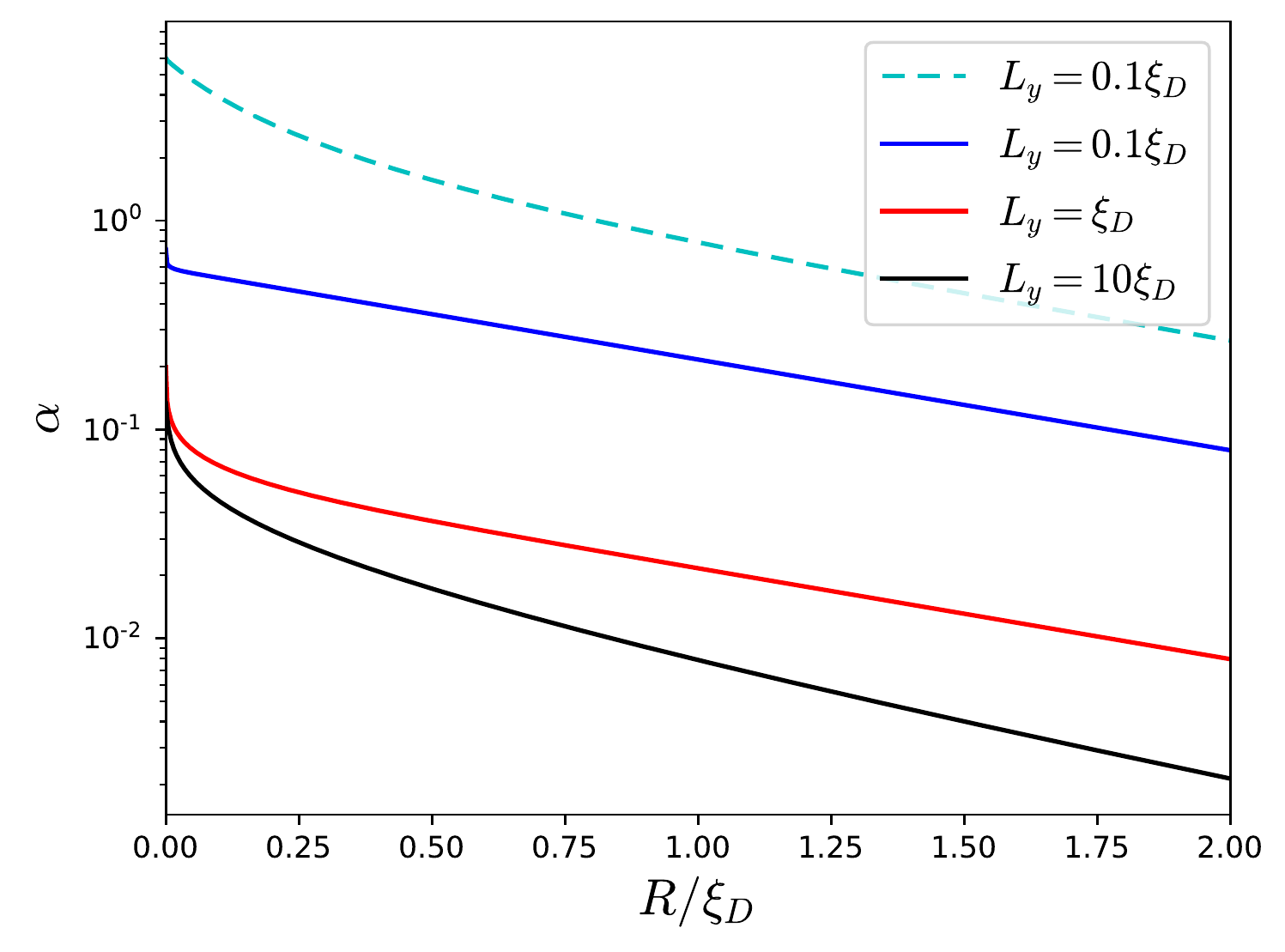}
    \caption{Dimensionless exchange coupling $\alpha$ as a function of the distance between dots $R$ in the absence of external magnetic field for different values of the  width $L_y$ of the superconductor. The solid lines are calculated for $L_z=100\xi_D$, and the dashed line for $L_z-R=0.2\xi_D$. All length scales are measured relative to the effective coherence length $\xi_D$ defined after \autoref{eq:alpha0}. For aluminum, we find the approximate value $\xi_D\approx 1\mu\text{m}$. The results are calculated numerically using \autoref{eq:asum3} and inserting values appropriate for aluminum discussed in \autoref{sec:conclusiones}.}
    \label{fig:dimensions}
\end{figure}

\begin{figure}[t]
  \includegraphics[width=\linewidth]{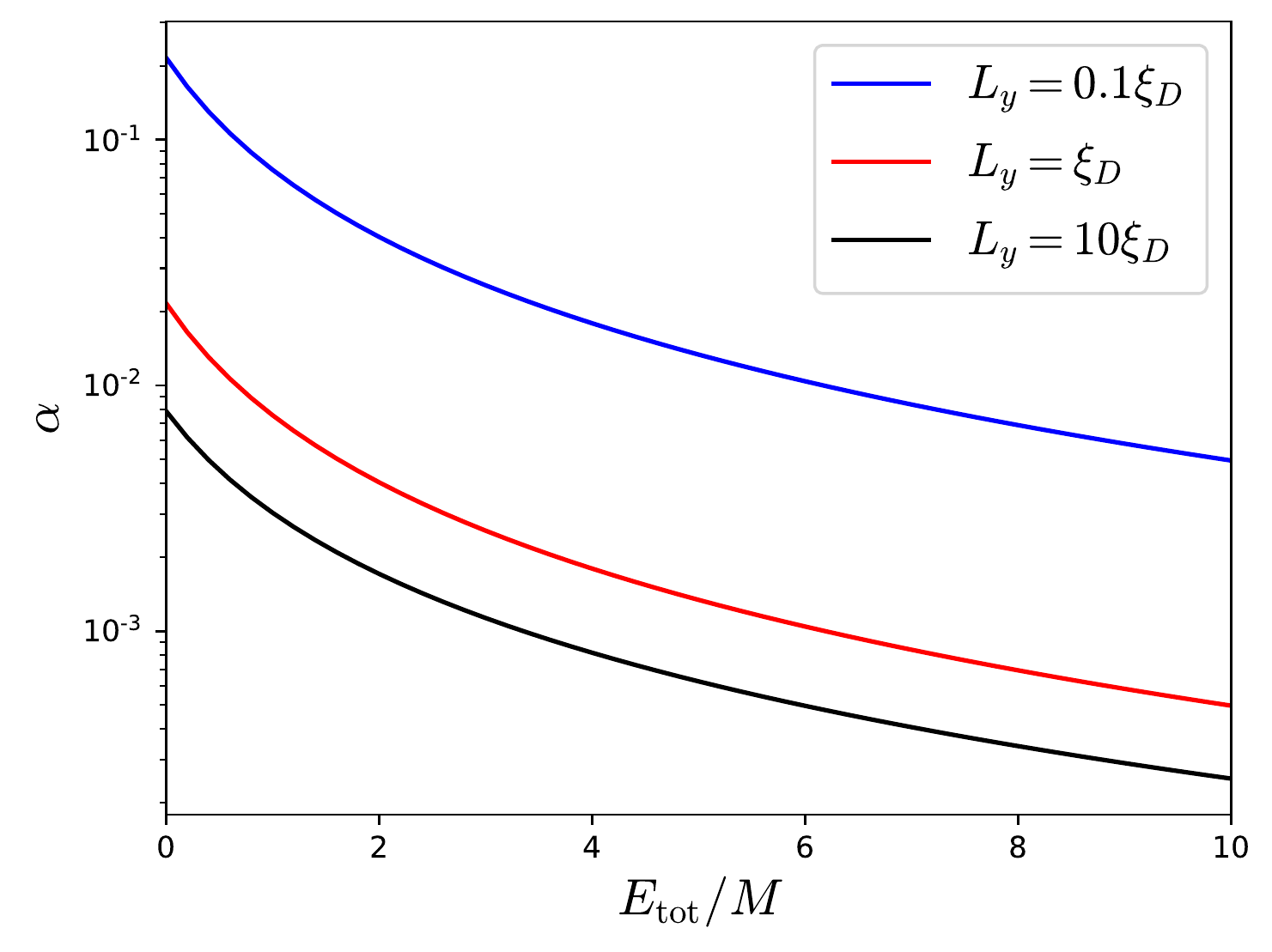}
    \caption{Dimensionless exchange coupling $\alpha$ as a function of the total Zeeman splitting $E_{\text{tot}}$ for an interdot distance of $R=\xi_D$ and with $\Delta_Z=0.1E_{\text{tot}}$ for three different values of $L_y$ and for $L_z=100\xi_D$. The approximate experimental values in the case for aluminum are $M\approx 1\,\mu\text{eV}$ and $\xi_D\approx 1\,\mu\text{m}$}
    \label{fig:dimetot}
\end{figure}

In the previous subsection, we have shown that the exchange interaction is reduced when an external magnetic field is applied. In this section, we will discuss how reducing the lateral dimension of the superconducting film ($L_y$) focuses the trajectories of the electrons and leads to an increase of $\alpha$. To this end, we have to include boundary effects. In Appendix~\ref{appendix:crossover}, we study the behavior of a disorder averaged pair of superconducting Green's functions below the gap as a function of the dimensions $L_y$ and $L_z$ of the superconductor (see \autoref{fig:setup}). We show that decreasing the size of the superconducting film increases the exchange interaction. Assuming $L_z> 2\xi_B$, the two-dimensional and quasi-one-dimensional limits are dictated by the width of the superconductor $L_y$ compared to the lengthscale $\xi_B$ defined in \autoref{eq:xis}. Indeed we find (assuming $M_B \gg 2\Delta_Z$ for simplicity)
\begin{equation}
    \alpha=\frac{\Delta}{2\pi M_B k_F \ell_e}
    \begin{cases}
    K_0(R/\xi_B) + 2K_0(L_z/\xi_B)  \\
    \qquad +K_0[(2L_z-R)/\xi_B],  & L_y \gg \xi_B, \\
    (\pi \xi_B/L_y)  (
    e^{-R/\xi_B} + \makebox[0pt][l]{$2e^{-L_z/\xi_B}$}  \\
    \qquad+e^{-(2L_z-R)/\xi_B} ), & L_y \ll \xi_B,
    \end{cases} 
\end{equation} 
where the result for $M_B \ll 2\Delta_Z$ can be obtained from \autoref{eq:alphacases}. We can see that the main difference between the two limits involves a factor $\xi_B/L_y$ as well as the change from a Macdonald function to a pure exponential decay. Asymptotically, these changes lead to an increase of $\alpha$ by a factor $(\xi_BR)^{1/2}/L_y$ when comparing the 1D to the 2D situation (see \autoref{fig:dimensions}). At the same time, the reflective boundaries in the $z$ direction also contribute positively to the exchange interaction. In particular, positioning the quantum dots close to the boundaries of the superconductor, $L_z-R\ll \xi_B$, can increase the exchange interaction $\alpha$ up to a factor of $4$. As a result, we find that geometric factors are crucial to optimize the exchange interaction. In particular, $\alpha$ can be increased by over an order of magnitude from its value in the two-dimensional limit by simply reducing the dimensions of the system so that $L_y \approx 0.1 \xi_D$, as can be seen in \autoref{fig:dimetot}, where the two-dimensional limit is depicted by the black line.

\section{\label{sec:conclusiones}Discussion and Conclusions}

We want to comment on potential experimental implications of our results. In particular, we would like to estimate the exchange interaction that can be achieved with the setup under discussion. Note that we estimate the strength of the exchange interaction through the dimensionless coupling parameter $\alpha$ [see \autoref{eq:alpha}], which gives the scaling of the exchange interaction in our setup with respect to the microscopic exchange interaction $J_0=\Gamma^2/\delta\varepsilon$, where we estimate $J_0\approx 10-100\,\text{GHz}$. Values of $\alpha$ of the order $10^{-2}-10^{-3}$ are therefore needed to ensure that the exchange interaction remains of the order of $100\,\text{MHz}$, which is a typical value for spin qubit operations \cite{BSOPLetal}. In order to obtain an improvement in scalability with this setup over current qubit architectures, we aim to achieve such exchange interaction strengths over interdot distances of $R\approx 1\,\mu\text{m}$.

The key elements that determine the strength of the exchange interaction are the effective coherence length in the absence of magnetic field $\xi_D$ [defined after \autoref{eq:alpha0}], the geometry of the superconductor, and the strength of the Zeeman splitting. Strictly speaking, the relevant length scale is $\xi_B$ and not $\xi_D$ [see \autoref{eq:xis}]; however, the effect of the Zeeman splitting on the coherence length is weak enough that we can assume $\xi_B\approx \xi_D$ for order-of-magnitude estimates. The effective coherence length dictates both the decay length of the exchange interaction as well as the dimensionality crossover. At $R\approx \xi_D$ the exchange interaction is not yet exponentially suppressed and a superconducting film of width $L_y \approx 0.1\xi_D$ is sufficient to reach the 1D limit. As we can see in \autoref{fig:dimensions}, achieving the 1D limit is a key objective for the optimal use of the setup, since it gives rise to an improvement of more than an order of magnitude over the 2D case. The importance of this improvement is highlighted when an external magnetic field is present, as large exchange interaction under the effect of large Zeeman splittings may only be achievable in the 1D limit (see \autoref{fig:dimetot}). 

We propose aluminum as a good candidate for the material of the superconducting film. Very thin aluminum films (thickness $L_x$ below 10\,nm) have a parallel critical field of a few Tesla and their order parameter is not significantly affected by the field up to a substantial fraction of the critical one~\cite{CWA}. However, the mean free-path $\ell_e$ is short, of the order of $L_x$. Since the parallel critical field scales as the inverse of the thickness, films of $L_x\simeq 30\,$nm should have both sufficiently high critical field (close to 1\,T) and sufficiently long mean free-path. Several times longer mean free path can be achieved by epitaxial growth~\cite{ANSC}, so we estimate $\ell_e\simeq 100\,$nm. With this mean free path, the value of Fermi velocity $v_F\simeq 2\times10^6\,$m/s$^2$, and the measured spin-orbit scattering rate~\cite{CWA,ANSC}, we arrive at $l_\text{so} \sim 1\,\mu$m~\footnote{We note that this is likely an underestimate of the actual spin-orbit length, since there is evidence that the sheet resistance decreases faster with thickness than one would expect from a linear increase of mean free path with $L_x$, and at the same time the spin-orbit scattering time increases with $L_x$~\cite{ANSC,LLWLL}}. Since, as discussed in Sec.~\ref{sec:spinorbit}, $l_\text{so}$ sets an upper limit for the effective coherence length $\xi_D$, we choose the detuning $M$ so that $\xi_D\sim 1\,\mu$m (while neglecting its possible decrease due to finite $l_\text{so}$). We therefore set $M= 5\times 10^{-3}\Delta\approx 1\,\mu\text{eV}$, where $\Delta \approx 200\,\mu\text{eV}$ is the superconducting gap of aluminum. Such a value for $M$ seems feasible, since the broadening of the peaks in the density of states for aluminum, as measured by the Dynes parameter $\gamma$, can be very small ($\gamma \approx 2\times 10^{-7}\Delta$~\cite{SKMP}). Finally, the prefactor $k_F \ell_e$ entering $\alpha$, see \text{e.g.} Eq.~(\ref{eq:alpha0}), can be rewritten as $2E_F \ell_e/v_F \simeq2\times 10^3$, where $E_F=11.6\,$eV is the Fermi energy in aluminum.

The Zeeman splitting depends both on the external magnetic field and the $g$-factors of the materials. For aluminum, a good estimate is $g^{\text{sc}}=2$ at low temperature and for the magnetic fields of interest, since in this case Fermi-liquid effects that renormalize the $g$-factor in the normal state are suppressed~\cite{CWA}. For the dots, $g^D$ can vary depending on several factors, mainly the material hosting the quantum dots and their shape, taking a wide range of values, from negative ones $g^D\approx-1$~\cite{BGYYetal,PF,SNKD} to values as large as $g^D\approx 50$~\cite{NNEBetal}, as well as values of the order the free electron one $g^D\approx 0.5$--$3$ \cite{KSSFetal}. We assume for simplicity $g^D\approx 2$, as appropriate for silicon~\cite{FKSNetal,ZTTHetal}, such that external magnetic fields of the order of $B\approx 100\,\text{mT}$ would lead to a total Zeeman splitting $E_{\text{tot}}\approx 10\,\mu\text{eV} \approx 10M$ \footnote{Note that in this case we have $\Delta_Z \ll M$ and $M_B \simeq M+E_\text{tot}$; hence it follows from Eq.~(\ref{eq:xis}) that $\xi_B/\xi_D \approx 0.55$. In the 1D regime for a long strip, this reduces $\alpha$ by a factor of 4, a reduction that can be counteracted by using a finite strip of length $L_z\simeq R$.}.
Looking at \autoref{fig:dimetot}, we estimate that using a small enough superconducting strip ($100\,\text{nm}$--$1\,\mu\text{m}$ width) as coupler between two quantum dots an exchange interaction in the order of $J\simeq 100\,\text{MHz}$ over distances of $R\simeq 1\,\mu\text{m}$ can be achieved. This shows that our setup can sustain an external magnetic field up to a few hundred mT and is therefore a viable approach to long-range coupling of spin qubits in a realistic setting.

\begin{acknowledgments}
This work was supported by the Deutsche Forschungsgemeinschaft (DFG) under Grant No. CA 1690/1.
\end{acknowledgments}

\appendix

\section{\label{appendix:angreen}Analytical calculation of disorder averaged product of Green's functions}

Performing the disorder average of a product of Green's functions does not simply equal to the product of the disorder averaged Green's functions. Instead, a self consistent equation to account for impurity scattering must be solved~\cite{AM}. In the case of superconductivity, the procedure is very similar to the normal state case, but the possibility of particle-hole conversion must be taken into account~\cite{AGD}. This can be done by working in Nambu space, as is thoroughly explained in Ref.~\cite{GRHC}. We do not outline the lengthy calculation in this appendix for simplicity, and point the reader to Ref.~\cite{GRHC} and Ref.~\cite{HCB}, which include all the necessary information to reproduce the calculation.

In momentum space, the disorder averaged product of Green's functions below the gap takes, in the diffusion approximation, the form
\begin{equation}
    \int \frac{d^dk}{(2\pi)^d}\overline{g(E,\bm{k})g(E',\bm{k}-\bm{q})}=\frac{c_0}{c_1+D^*q^2}
    \label{eq:kspace}
\end{equation}
where $D^*$ is the effective diffusion constant and $c_0,c_1>0$ are parameters dependent on $E,\,E'<\Delta$. The analytical expressions for $c_0$ and $c_1$ are not trivial in the general case where $E\neq E'$. We therefore obtain analytical results in two limits of interest by doing a series expansion to disregard negligible terms. We define
\begin{equation}
    D_{\omega}\equiv \int \frac{d^dk}{(2\pi)^d}\overline{g(\Delta-m,\bm{k})g(\Delta-m-\omega,\bm{k}-\bm{q})}.
\end{equation}
The dimensionless coupling constant $\alpha$ can be obtained from $D_{\omega}$ after performing a Fourier transform into real space and substituting the appropriate values for $m$ and $\omega$ [$m\equiv M_B$ and $\omega \equiv 2\Delta_Z$ for \autoref{eq:alphatuned}]. We calculate the analytical expression of $D_{\omega}$ in the limits $\omega \ll m\ll\Delta$ and $m\ll \omega \ll \Delta$. In both cases we work in the disordered limit, \textit{i.e.}, $\ell_e\Delta \ll v_F$.

In the limit $\omega \ll m\ll\Delta$, we obtain
\begin{equation}
\begin{split}
D_{\omega}=\frac{\pi \rho_0 \Delta}{2m}\frac{1}{\sqrt{8 m \Delta}+ \sqrt{\frac{9 \Delta }{2 m}} \omega\, +D(1+\frac{\omega}{2m}) q^2 }
    \end{split}
    \label{eq:smallfield}
\end{equation}
with the diffusion constant $D=v_F \ell_e/d$ and where the result for $\omega=0$ and $m=M$ corresponds to the case without magnetic field considered in Ref.~\cite{HCB}. In the limit $m\ll \omega \ll\Delta$ we have
\begin{equation}
\begin{split}
D_{\omega}=\frac{ \pi \rho_0\Delta }{2\sqrt{\omega m}}\frac{1}{ \sqrt{2\omega  \Delta}+\sqrt{2\Delta m}+D
q^2}.
    \end{split}
    \label{eq:bigfield}
\end{equation}
After Fourier transforming into two dimensional real space, we have, in the $\omega \ll m \ll \Delta$ limit,
\begin{equation}
\begin{split}
D_{\omega}=\frac{ \rho_0 \Delta}{4m D}K_0(R/\xi_B)\left(1-\frac{\omega}{2m}\right);
    \end{split}
    \label{eq:smallfieldr}
\end{equation}
whereas for $m \ll \omega \ll\Delta$, we find
\begin{equation}
\begin{split}
D_{\omega}=\frac{\Delta \rho_0}{4\sqrt{\omega m} D}K_0(R/\xi_B).
    \end{split}
    \label{eq:bigfieldr}
\end{equation}
Here, the coherence length is given by 
\begin{equation}
\xi_B=
\begin{cases}
\frac{D^{1/2}}{(8\Delta m)^{1/4}}(1-\frac{\omega}{8m}),& \omega \ll m \ll\Delta,\\
\frac{D^{1/2}}{(2\Delta \omega)^{1/4}}, & m \ll \omega \ll\Delta .
\end{cases}
    \label{eq:decayrates}
\end{equation}

\section{\label{appendix:heff}Calculation of the effective Hamiltonians}

We treat the tunneling Hamiltonian $H_T$ [see \autoref{eq:htunnel}] as a perturbation in order to construct an effective Hamiltonian for the states $\{ \ket{\uparrow,\downarrow},\ket{\downarrow,\uparrow}\}$ using the Schrieffer-Wolff transformation~\cite{BDL}. In order to do so, we divide the Hilbert space in two groups of states well separated in energy, which we conventionally denote as `high' and `low' states. The low states consist of the subspace of interest, the exchange subspace $\mathcal{E}=\{ \ket{\uparrow,\downarrow},\ket{\downarrow,\uparrow}\}$. The high states consist of the states with one electron in each dot $\ket{\sigma,\sigma}$, the intermediate virtual states with one electron in the superconductor and one in the second dot $ \ket{\bm{k}\sigma;0,\sigma'}$, and the state with both electrons in the right dot $\ket{0,\uparrow\downarrow}=d_{2\uparrow}^{\dag}d_{2\downarrow}^{\dag}\ket{0,0}$. We work in the basis $\{\ket{\uparrow,\downarrow}$, $\ket{\downarrow,\uparrow}$, $\ket{\uparrow,\uparrow}$, $\ket{\downarrow,\downarrow}$, $\ket{\bm{k}\uparrow;0,\downarrow}$, $\ket{\bm{k}\downarrow;0,\uparrow}$, $\ket{\bm{k}\uparrow;0,\uparrow}$, $\ket{\bm{k}\downarrow;0,\downarrow}$, $\ket{0,\uparrow\downarrow}\}$, which we will enumerate as $\ket{n}$ with $n$ from 1 to 9 in the order shown above for simplicity. Note that we do not take into account the states where one electron resides in the first dot and another one in the superconductor. This is due to the fact that the dots are tuned such that $\varepsilon_{1,1}-\varepsilon_{1,0}\ll \varepsilon_{1,1}-\varepsilon_{0,1}$; therefore a single electron residing in the second dot and being transferred to the superconductor starts from an energy much lower than the gap and its probability of virtual propagation to the first dot will be negligible in comparison to that of an electron moving from the first to the second dot (see \autoref{fig:energies}). For a given momentum $\bm{k}$, the Hamiltonian of the system projected onto our basis has the form
\begin{align}
    H_{\bm{k}}= \begin{pmatrix}
       E_1 & 0 & 0 & 0 & V_{1,5} & 0 & 0 & V_{1,8} & 0 \\
     0 & E_2 & 0 & 0 & 0 & V_{2,6} & V_{2,7} & 0 & 0 \\
     0 & 0 & E_3  & 0 & 0 & V_{3,6} & V_{3,7} & 0 & 0 \\
     0 & 0 & 0 & E_4  & V_{4,5} & 0& 0 & V_{4,8} & 0 \\
     V_{5,1} & 0 &0 & V_{5,4} & E_5 & 0 & 0 & 0 & V_{5,9} \\
     0 & V_{6,2}& V_{6,3} & 0 & 0 & E_6 & 0 & 0 & V_{6,9} \\
    0 & V_{7,2} & V_{7,3} & 0 & 0 & 0 & E_7 & 0 & V_{7,9} \\
    V_{8,1} & 0& 0 & V_{8,4} & 0 & 0 & 0 & E_8 & V_{8,9} \\
    0 & 0& 0 & 0 & V_{9,5} & V_{9,6} & V_{9,7} & V_{9,8} & E_9
    \end{pmatrix}\!,
    \label{eq:htotalm}
\end{align}
where the values of the energies in the diagonal can be obtained using \autoref{eq:hdot}, \autoref{eq:hbcs} and \autoref{eq:hz}, and the off-diagonal coupling terms originate from the tunneling Hamiltonian [see \autoref{eq:htunnel}].
We calculate the perturbation theory terms contributing to the effective Hamiltonian in the subspace spanned by $\{\ket{\uparrow,\downarrow},\ket{\downarrow,\uparrow}\}$ up to fourth order: this is the lowest order contributing to the exchange interaction, which is mediated by virtual transitions through the state $\ket{0,\uparrow\downarrow}$. The first and third order terms are zero. For the zeroth and second order term we have, respectively,
\begin{equation}
        H^{(0)}=\sum_{\bm{k}}\sum_{i\in \{1,2\}} E_i \ket{i}\bra{i}
       \label{eq:h0}
\end{equation}
\begin{equation}
        H^{(2)}=\frac{1}{2}\sum_{\bm{k}}\sum_{i,i',j} V_{i,j}V_{j,i'}\Bigl(\frac{1}{E_i-E_j}+\frac{1}{E_{i'}-E_j}\Bigr)\ket{i}\bra{i'}
    \label{eq:h2}
\end{equation}
with $i,i' \in \{1,2\}$ and $j \in \{3,\dots, 9\}$. (Note that in reality only terms with $j \in \{ 5,6,7,8\}$ give a non-zero contribution at this order). 

We are only interested in fourth order perturbation terms that affect the exchange interaction $J$ [see \autoref{eq:hqubit}]. Other fourth order terms will be negligible in comparison to $H^{(2)}$. We therefore do not calculate all fourth order perturbation terms, but instead restrict our attention to the term
\begin{align}
    H^{(4)}_\text{ex}&=\frac{1}{2}\sum_{\bm{k}}\sum_{i,i',j,j',j''}V_{i,j}V_{j,j''}V_{j'',j'}V_{j',i'} \nonumber \\&\quad\Bigl[\frac{1}{(E_i-E_{j''})(E_i-E_j)(E_i-E_{j'})}\\&\qquad+\frac{1}{(E_{i'}-E_{j''})(E_{i'}-E_j)(E_{i'}-E_{j'})}\Bigr]\,|i\rangle\langle i'| \nonumber
\end{align}
as it is the only fourth order term that can include the state $\ket{0,\uparrow\downarrow}$, and as such, any contribution to the exchange interaction is included in it. 

Introducing the explicit values for $E_i$ and $V_{i,j}$ with $i,j \in \{1,\dots,9\}$ as given by the Hamiltonians defined in \autoref{sec:setup}, and performing a local rotation in the $xy$-plane such that the coupling between $\ket{\uparrow,\downarrow}$ and $\ket{\downarrow,\uparrow}$ is entirely real (that is, to ensure there is no $\tau_y$ term in $H_{\mathcal{E}}$) we obtain the explicit expressions for the exchange Hamiltonian terms [\autoref{eq:hqubit}]. For the case where all processes are virtual, \textit{i.e.}, $\epsilon_1+|E_{\text{tot}}|,\epsilon_1+|\Delta_Z|<\Delta$, they are given by
\begin{equation}
    \begin{split}
    \beta&=\frac{1}{2}\Bigl[t_1^2\big(\overline{g(\epsilon_1+\Delta_Z,0)}-\overline{g(\epsilon_1-\Delta_Z,0)}\big)\\&
    \quad+|t_{f1}|^2\big(\overline{g(\epsilon_1+E_{\text{tot}},0)}-\overline{g(\epsilon_1-E_{\text{tot}},0)}\big)\Bigr],
         \label{eq:bz}
    \end{split}
\end{equation}
and
\begin{equation}
    \begin{split}
        J&=\frac{2}{\delta\varepsilon}\Bigl[t_1^2t_2^2\overline{g(\epsilon_1+\Delta_Z,R)g(\epsilon_1-\Delta_Z,R)}\\&
        \quad+|t_{f1}|^2|t_{f2}|^2\overline{g(\epsilon_1-E_{\text{tot}},R)g(\epsilon_1+E_{\text{tot}},R)}\\&
        \quad+t_1t_2|t_{f1}||t_{f2}|\big(\overline{g(\epsilon_1-E_{\text{tot}},R)g(\epsilon_1+\Delta_Z,R)}\\&
        \quad+\overline{g(\epsilon_1+E_{\text{tot}},R)g(\epsilon_1-\Delta_Z,R)}\big)\Bigr].
         \label{eq:bx}
    \end{split}
\end{equation}

\section{\label{appendix:crossover}Dimensional crossover}

In the original proposal of Ref.~\cite{HCB}, the superconducting coupler was taken for simplicity to be an infinite, two-dimensional film. However, one can expect that reducing the size of the superconducting coupler will enhance the probability of virtual transition from one dot to the other, and hence the exchange interaction, since the probability of propagation between the two dots increases when the motion of the electrons is confined. In the following, we study the behavior of $\alpha$ in a setup with a finite sized superconductor of lengths $L_y$ and $L_z$ in the $y$ and $z$ direction respectively. We show that the relevant length scale that determines dimensionality is $\xi_B$ [see \autoref{eq:xis}], where the maximum value $\xi_B=\xi_D$ is obtained at zero external magnetic field.

We assume both quantum dots to be placed at positions $(R/2,0)$ and $(-R/2,0)$ respectively, with the superconductor defined between $-L_z/2$ and $L_z/2$ in the $z$ direction and between $-L_y/2$ and $L_y/2$ in the $y$ direction (note that we are using $y$ and $z$ coordinates for the relevant direction). The dimensionless coupling parameter, as can be seen in \autoref{eq:alpha}, takes the form $\alpha=\sum_i \eta_i \alpha_i$ where $\eta_i$ is a coefficient made of four tunneling amplitudes and $\alpha_i$ is a disorder averaged product of Green's functions. To study the dimensional crossover of the system we focus on understanding the behavior of each $\alpha_i$ for different limits of $L_y$. In the diffusive limit in momentum space, each $\alpha_i$ takes the general form in \autoref{eq:kspace}.

The relative momentum between Green's functions $q^2=|q_y|^2+|q_z|^2$ is discretized due to Neumann boundary conditions  in both directions. In particular, we have $q_y=\frac{n_y\pi}{L_y}$ and $q_z=\frac{n_z\pi}{L_z}$ with $n_y,n_z\in \mathbb{Z}$. Then each $\alpha_i$ can be represented as the Fourier series~\cite{AM},
\begin{align}
    & \alpha_i = \nonumber\\ & \quad \frac{A}{L_z L_y }\sum_{n_z,n_y}\frac{\text{cos}^2(n_y\pi/2)\text{cos}(z_+n_z\pi/L_z)\text{cos}(z_-n_z\pi/L_z)}{c_{1}+D^*(q_y^2+q_z^2)}\nonumber\\&=\frac{A}{L_z L_y }\sum_{n_z,n_y}\frac{\text{cos}(z_+n_z\pi/L_z)\text{cos}(z_-n_z\pi/L_z)}{c_{1}+D^*(4 q_y^2+q_z^2)},
    \label{eq:asum1}
\end{align}%
with $A=c_0/2\pi^2\rho_0^2$ and where $z_+=(L_z+R)/2$ and $z_-=(L_z-R)/2$. This can be rewritten as
\begin{equation}
    \alpha_i=\frac{A}{2L_z L_y }\sum_{n_z,n_y}\frac{e^{in_z \pi }+e^{in_z \pi R/L_z }}{c_{1}+D^*(4 q_y^2+q_z^2)}.
    \label{eq:asum2}
\end{equation}%
Substituting the values of $q_y$ and $q_z$ and evaluating the sum over $n_z$ (this can be done with the help of Poisson's summation formula), we obtain
\begin{equation}
\begin{split}
    \alpha_i&=\frac{A}{2c_{1}\xi_{B} L_y }\sum_{n_y}\frac{1}{\sqrt{1+(2\pi n_y \xi_{B}/L_y)^2}}\\
    &\quad\times \frac{1+\cosh\left({\frac{L_z-R}{\xi_{B}}\sqrt{1+(2\pi n_y \xi_{B}/L_y)^2}}\right)}{\sinh\left(\frac{L_z}{\xi_{B}}\sqrt{1+(2\pi n_y \xi_{B}/L_y)^2}\right)},
    \label{eq:asum3}
    \end{split}
\end{equation}%
where we have introduced  $\xi_{B}=\sqrt{D^*/c_{1}}$.
We can obtain approximate expressions for this equation in different limits. We first assume $L_z > 2\xi_{B}$, such that we can approximate $\sinh(x)\approx \tfrac12 e^{x}$. The expression in \autoref{eq:asum3} then becomes
\begin{align}
    \alpha_i&=\frac{A}{2c_{1}\xi_{B} L_y }\sum_{n_y}\frac{1}{\sqrt{1+(2\pi n_y \xi_{B}/L_y)^2}}
 \nonumber\\
    &\quad\times\left[ e^{-R/\xi_{B} \sqrt{1+(2\pi n_y \xi_{B}/L_y)^2}} \right.\nonumber  \\& \left. \quad+2e^{-L_z/\xi_{B}\sqrt{1+(2\pi n_y \xi_{B}/L_y)^2}}\right. \nonumber  \\& \left. \quad+e^{-(2L_z-R)\sqrt{1+(2\pi n_y \xi_{B}/L_y)^2}}\right], 
    \label{eq:asum4}
\end{align}%
we can now evaluate this expression in the different limits for $2\pi \xi_{B}/L_y$. For $2\pi \xi_{B}/L_y \ll 1$, we find the two-dimensional limit, where we can turn the sum into an integral and obtain
\begin{equation}
\begin{split}
    \alpha_i&\approx \frac{A}{2c_{1} \pi \xi_{B}^2}\left[ K_0(R/\xi_{B}) + 2K_0(L_z/\xi_{B}) \right. \\ & \left. \qquad+K_0\big((2L_z-R)/\xi_{B}\big) \right].
    \label{eq:asum5}
    \end{split}
\end{equation}%
In the one-dimensional limit $2\pi \xi_{B}/L_y \gg 1$ we can instead approximate
\begin{align}\label{eq:asum6}
    \alpha_i&\approx \frac{A}{2c_{1}\xi_{B} L_y }\Big[ e^{-R/\xi_{B}} +  2e^{-L_z/\xi_{B}}  + e^{-(2L_z-R)/\xi_{B}}
    \nonumber\\ & \quad+ \sum_{n_y\geq 1} \frac{L_y }{\pi n_y \xi_{B}} \Bigl( e^{-2 \pi n_y R/L_y} +  2e^{-2 \pi n_y  L_z/L_y} \nonumber \\
    &\quad+ e^{-2 \pi n_y (2L_z-R)/L_y} \Bigr) \Big], 
\end{align}%
where the sum can be evaluated using $\sum_{n\geq 1} e^{-a n}/n=-\log (1-e^{-a})$.

If a magnetic field is present, each $\alpha_i$ is characterized by a different length scale $\xi_{B}$. A strict approach would then require $L_y$ to be smaller (larger) than all of those length scales to enter the 1D (2D) limit. In reality, however, due to the fact that $|t_f|\ll t$, the main contribution to the exchange interaction is given by \autoref{eq:alphatuned}, such that the decay length defining the dimensionality crossover will be given by \autoref{eq:xis}.

\bibliography{refs}

\end{document}